\documentclass[twocolumn]{IEEEtran} 
\usepackage{graphicx}

\def\BibTeX{{\rm B\kern-.05em{\sc i\kern-.025em b}\kern-.08em
  T\kern-.1667em\lower.7ex\hbox{E}\kern-.125emX}}

\newcommand{\BaBar}{\textsc{BaBar}}

\begin{document}

\title{Design and Performance of the Level 1
Calorimeter Trigger for the \BaBar\ Detector
\thanks{Manuscript first received November 2, 2000.
This work was funded by the Particle Physics and Astronomy Research
Council, UK.}}

\author{
P.~ D.~ Dauncey,\thanks{Paul Dauncey and David Price are with the
Department of Physics, Imperial College, University of London,
Prince~Consort~Road, London SW7~2BW,~UK. For inquiries about this
paper, contact Paul Dauncey; telephone: +44-20-7594-7803,
email: P.Dauncey@ic.ac.uk.}
J.~ C.~ Andress,\thanks{John Andress, Nicole Chevalier, Neil Dyce, Brian
Foster, Alexander Mass, Jason McFall, Steven Nash, Uli Sch\"afer
and David Wallom are with the 
Department of Physics, University of Bristol, Tyndall Avenue,
Bristol, BS8~1TL,~UK.}
T.~ J.~ Adye,\thanks{Timothy Adye, Brian Claxton and Senerath Galagedera
are with the Rutherford Appleton Laboratory,
Chilton, Didcot, OX11~0QX,~UK.}
N.~ I.~ Chevalier,
B.~ J.~ Claxton,
N.~ Dyce,
B.~ Foster,
S.~ Galagedera,
A.~ Kurup,\thanks{Ajit Kurup, Paul McGrath and Iain Scott are with the
Physics Department, Royal Holloway and Bedford New College,
University~of~London, Egham, TW20~0EX,~UK.}
A.~ Mass,
J.~ D.~ McFall,
P.~ McGrath,
S.~ J.~ Nash,
D.~ R.~ Price,
U.~ Sch\"afer,
I.~ Scott
and
D.~ C.~ H.~ Wallom
}

\maketitle

\begin{abstract}
Since May 1999 the \BaBar\ detector has been taking data at the PEP-II
asymmetric electron-positron collider at the Stanford Linear Accelerator
Center, California. This experiment requires a very large data sample
and the PEP-II accelerator uses intense beams to deliver the high
collision rates needed. This poses a severe challenge to the
\BaBar\ trigger system, which must reject the large rate of background
signals resulting from the high beam currents whilst accepting the
collisions of interest with very high efficiency.
One of the systems that performs this task is the Level 1
Calorimeter Trigger, which identifies
energy deposits left by particles in the \BaBar\ calorimeter. It is a
digital, custom, fixed latency system which makes heavy use of
high-speed FPGA devices to allow flexibility in the choice of data filtering
algorithms. Results from several intermediate processing stages are read out,
allowing the selection algorithm to be fully analysed and optimized offline.
In addition, the trigger is monitored in real time by
sampling these data and cross-checking each stage of the
trigger calculation against a software model. The design, implementation,
construction and performance of the Level 1 Calorimeter Trigger during the
first year of \BaBar\ operation are presented.
\end{abstract}


\section{PEP-II and \BaBar}
\PARstart{T}{he} \BaBar\ experiment~\cite{ref:tdr}, \cite{ref:osaka}
has been built to study $CP$-violation in the decays of
$B$ mesons. These are produced in the decays of $\Upsilon(4S)$ mesons
created in $e^+e^-$ collisions at the PEP-II collider at the Stanford
Linear Accelerator Center in California, USA.
$CP$-violating effects are expected to be subtle, and so these measurements 
require very large statistical samples of $B$ mesons.
This means that the PEP-II machine must deliver unprecedentedly
high luminosity, and indeed it has already set a new world
record luminosity of greater than $2\times 10^{33}$ cm$^{-2}\,$s$^{-1}$
during the first year of operation, enabling \BaBar\ to record
a data sample of approximately 20 million $B$ meson decays.

The high luminosity of PEP-II has been obtained by using very
high beam currents, of order 1 A. The PEP-II machine design achieves this
by filling the machine
with 1658 bunches in each beam, with a bunch collision rate of 238 MHz.
A serious consequence of this high current is that
the machine-induced backgrounds have been very significant. This has provided
a major challenge for the \BaBar\ trigger and data acquisition system,
as this unwanted background must be rejected without losing the 
$B$ meson decays of interest.
Since high statistics are vital for this experiment,
the trigger design emphasized the 
need for high efficiency.

\section{The Level 1 Trigger Requirements}
The \BaBar\ trigger has two levels, a hardware Level~1 and a software Level~3
trigger.
This article describes one of the two main sections of the Level~1 trigger,
the ElectroMagnetic calorimeter Trigger (EMT). The other main section,
the Drift Chamber Trigger (DCT), has been described
previously~\cite{ref:dct}. A third component, the Instrumented Flux
return Trigger (IFT) is used to collect
events with muon particles which are used for calibration purposes. This
part of the trigger is not primarily designed to be efficient for $B$
mesons.

These two main subsystems work independently and in parallel, processing data
from the
calorimeter and drift chamber detectors respectively, and seek 
characteristics of particles coming from $B$ meson decays
to identify signals.
These signals, termed trigger ``primitives'', are then passed
to a third system,
the Global Level~1 Trigger (GLT). Here the individual primitives
are combined to form
a picture of the whole particle collision, and the GLT then decides
whether or not to accept this collision. This trigger decision is
passed to the fast control system where it is distributed to the
whole detector.
Fig.~1
shows the systems with which the Level~1 trigger interacts.
\begin{figure}[htbp]
\label{figure:L1TRIG}
\centerline{\includegraphics[width=3.4in,clip]{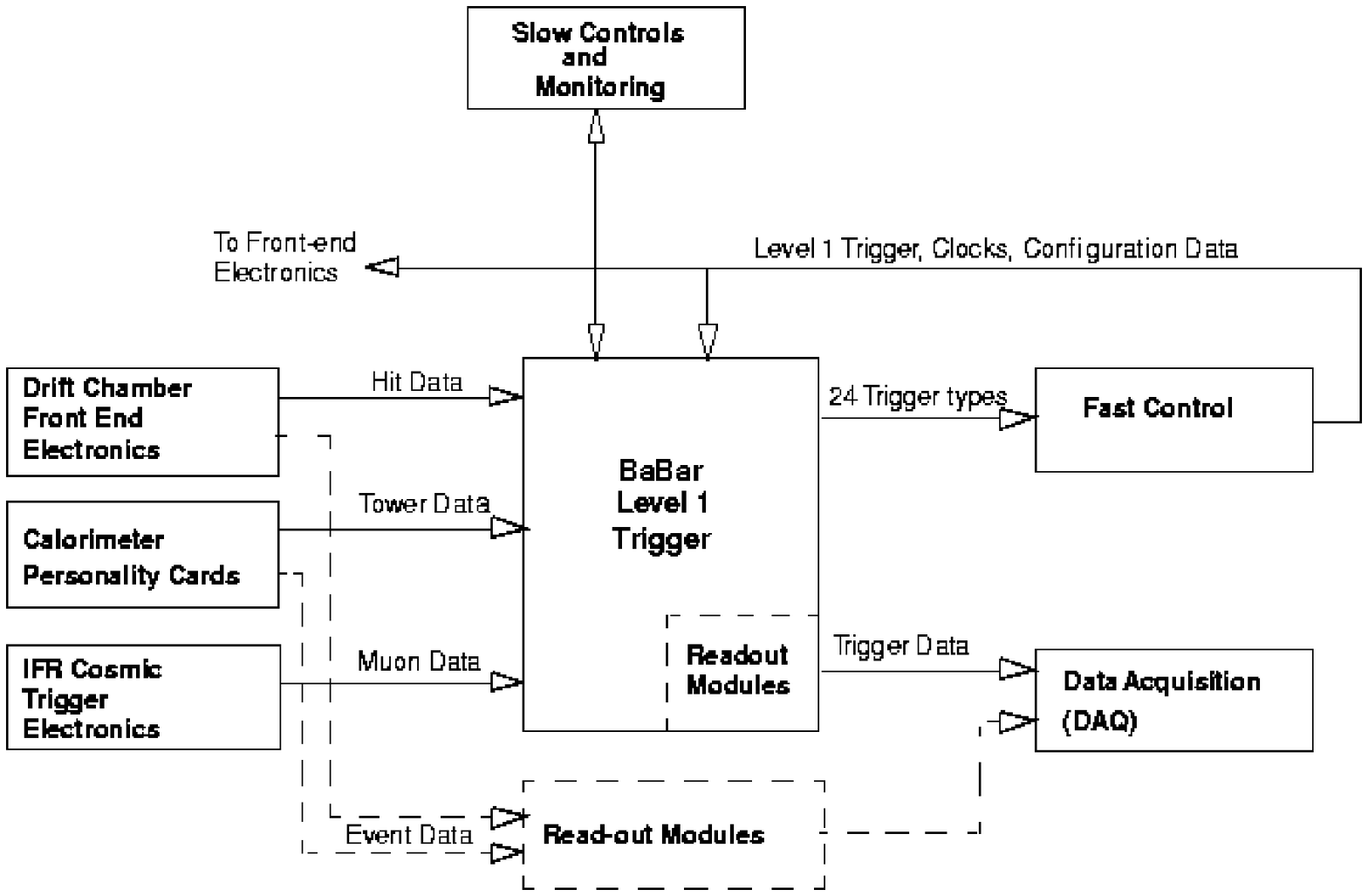}}
\caption{Context of the Level~1 trigger within the \BaBar\ experiment.
The EMT, together with the other parts of the Level~1 trigger, is
contained within the central box labelled ``BaBar Level~1 Trigger''.}
\end{figure}

The technical requirements for the Level~1 trigger system as a whole
are set by the
data acquisition~\cite{ref:daq}
and other electronics systems of \BaBar. To ensure maximum
redundancy, the EMT and DCT subsystems were each designed to satisfy
these requirements independently.

All the front-end electronics systems of the 
\BaBar\ detector buffer their data
for 12 $\mu$s, which defines the latency allowed for the Level~1
trigger decision.
The collision rate of 238 MHz is effectively continuous as far as the
time resolution of the drift chamber and calorimeter detectors are
concerned, and so the Level~1 trigger has the task of 
determining the time of the collision of interest to within 1 $\mu$s.
Each system in \BaBar\ then reads out data in a window
of at least 1 $\mu$s, in order to ensure that the data from the
triggered collision are fully collected.

The data acquisition system
is designed to accept
a maximum Level~1 output rate of
2~kHz. This is much higher than the rate of physics collisions, which
totals around 120~Hz. The Level~1 trigger is designed to be as 
unbiased as this maximum output rate allows, and to accept all
physics events without preselection. The challenge is to provide such
a trigger whilst still rejecting sufficient background to
keep the total rate below 2~kHz.

A number of other requirements must be satisfied in addition to high
efficiency. The efficiency must be measurable accurately, the trigger
must be stable and robust to varying background levels, and it must be
flexible enough to be adaptable to unexpected operating conditions.
The first of these is ensured by the redundancy of
the EMT and DCT triggers. The remainder were satisfied by designing
the trigger to be able to operate at up
to ten times the expected level of background. 
The EMT design described below has been shown 
to fulfill all the requirements.

\section{The EMT Data Interfaces and Algorithm}
The EMT receives its data from the electromagnetic calorimeter.
There are 6580 channels in the electromagnetic calorimeter, each giving
a 16-bit energy value at 3.7 MHz. These data are extracted continuously
(without triggering) 
from the front-end electronics via 280
fibre-optic links~\cite{ref:glink}.
In order to reduce the EMT input data rate, the data are summed 
over typically
24 channels before being sent to the EMT, since a physical energy 
deposit is usually spread over multiple calorimeter channels.
This reduces the input rate to 2.1 GBytes/s, which are transmitted 
as serial 59.5 MHz differential ECL signals. To compensate for 
differing cable lengths and component time delays, the data are
resynchronised to the 59.5 MHz clock when they are received by the EMT.

The technique used by the EMT to identify particle energy deposits 
relies on the fact that the probability of multiple particles arriving close 
together in time in the same region of the calorimeter is low.
This is true even in the high background conditions of PEP-II.
This means energy deposits may be summed over azimuthal ($\phi$) slices 
of the calorimeter, yielding 40 so-called ``$\phi$ strips'', 
without significantly increasing the probability of pile-up.
This considerably simplifies the trigger logic, 
and the EMT processing now becomes a one-dimensional,
rather than two-dimensional, problem. 
Each $\phi$ strip corresponds to
165 crystals; these are added in overlapping pairs to form 40 ``$\phi$ sums''
in order to fully contain energy deposits which extend into a neighbouring
$\phi$ region.

An overview of the algorithm
applied to the $\phi$ sums is shown in Fig.~2.
\smallskip
\begin{figure}[htbp]
\label{figure:phisum}
\centerline{\includegraphics[width=3.4in,clip]{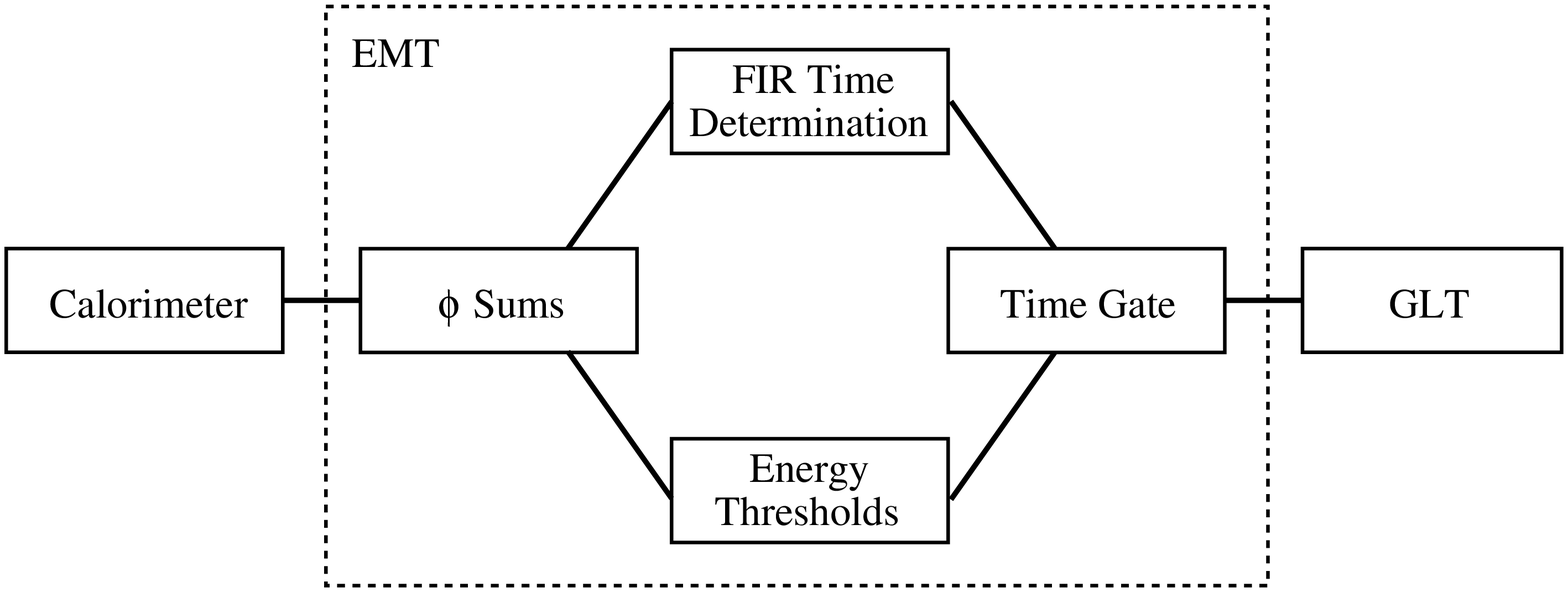}}
\caption{Simplified diagram of the stages of the EMT trigger algorithm.}
\end{figure}
The data from each sum are used in two ways: to find the
time at which the energy was deposited, and to compare
the total energy against several
thresholds.

The time is determined using a digital 8-tap finite impulse response
(FIR) filter which operates on the summed energy signal from each 
$\phi$ sum at a frequency of 3.7 MHz. 
Before digitisation, the calorimeter signal is
shaped by a three stage differentiator-integrator-integrator (CR-RC-RC)
circuit with time constants of 0.8 $\mu$s, 0.25 $\mu$s, 0.25 $\mu$s
respectively.
The EMT FIR weights
are chosen such that this shaped signal drives
the filter output from positive to negative at a fixed time interval
after the particle arrives.
This is illustrated in Fig.~3.
It is this ``zero-crossing'' which is
detected and used as the basic time estimate.
To increase the accuracy of this
determination, the EMT also performs 
a linear interpolation between consecutive FIR output values to approximate the
output at 7.4 MHz. 
\begin{figure}[htbp]
\label{figure:shape}
\centerline{\includegraphics[width=3.4in,clip]{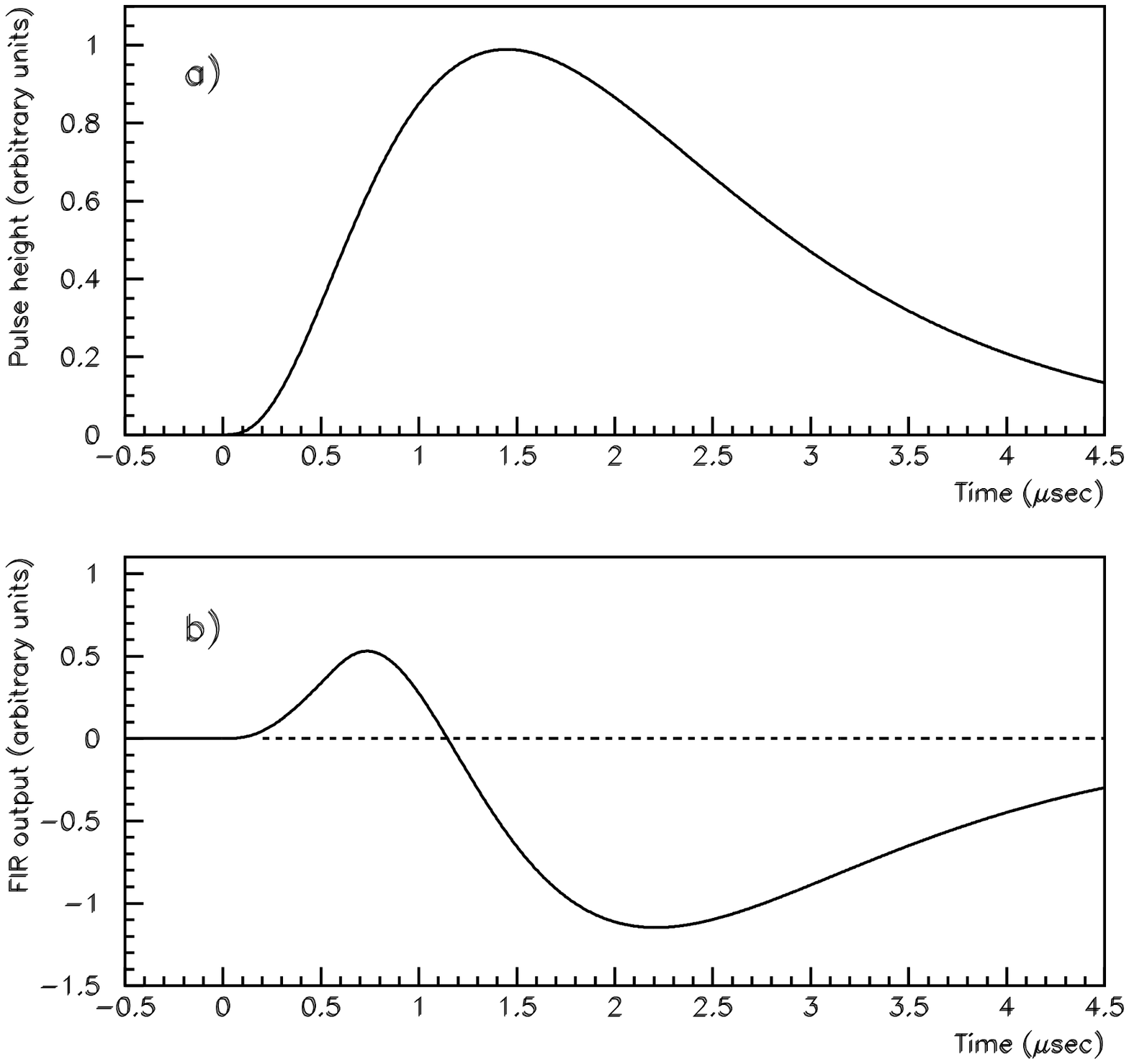}}
\caption{Illustration of an ideal pulse of a) the shaped calorimeter
signal input to the EMT and b) the corresponding FIR output. The latter
is shown for the FIR weights actually used in the first year of data-taking,
which were $+1$ and $-2$ for the first and third weights,
and zero for the remaining six.
This choice gives a zero-crossing approximately 1.2 $\mu$s after the time
of the energy deposit.}
\end{figure}

In parallel, the energy values of the $\phi$ sums are
compared against three threshold values. These configurable thresholds
are typically set to values around 120~MeV, 300~MeV and 800~MeV.
The first is designed to be efficient for minimum ionising particles,
which on average deposit 180~MeV in the calorimeter.
Typical GLT triggers would require four non-neighbouring
$\phi$ sums above this
lowest threshold, or three $\phi$ sums with a minimum $\phi$ angle
separation between at least one pair of them of $120^\circ$. 
The other thresholds are
set as low as possible, consistent with
reasonable background trigger rates.
These data are typically used in the GLT for
triggers requiring two $\phi$ sums, such as
two $\phi$ sums above the 300 MeV threshold with
a minimum $\phi$ angle separation of $120^\circ$, or two 
non-neighbouring $\phi$ sums
above the 800 MeV threshold with no further angle requirement.

Each of the three thresholds from each $\phi$ sum 
has a corresponding output bit. The bits from pairs of $\phi$ sums are
put through an OR so as to reduce the number of bits for the whole
range of $\phi$ from 40 to 20; this is necessary as the GLT does not have
the input bandwidth for a 40-bit wide array. The degradation on the
trigger performance because of this OR has been checked to be negligible.
These bits together form the EMT primitives. 
The number and position of these bits are used in the GLT to
determine if a valid trigger condition existed, such as the ones
outlined above.
The time determination from the FIR filter is used to gate the
threshold bits so that they are only set on around the time of a
valid energy deposit.
The bits are transmitted to the GLT 
as differential PECL at a rate of 7.4 MHz, giving a total data rate
for these data of 450 Mbits/s.

\section{Implementation of the EMT}
The algorithm outlined above was implemented so that data from
each $\phi$ sum are processed by a single ``Algorithm''
Xilinx 4020E FPGA clocked at 59.5 MHz.
All 40 such components operate in parallel. They are distributed over
ten Trigger Processor Boards (TPB), each containing four
such FPGA's, all with identical firmware. Each FPGA, and TPB,
runs independently of the others, so the data cannot be considered
at a ``global'' event level until they reach the GLT.
An overview of the whole EMT system is shown in Fig.~4.
\medskip
\begin{figure}[htbp]
\label{figure:system}
\centerline{\includegraphics[width=3.4in,clip]{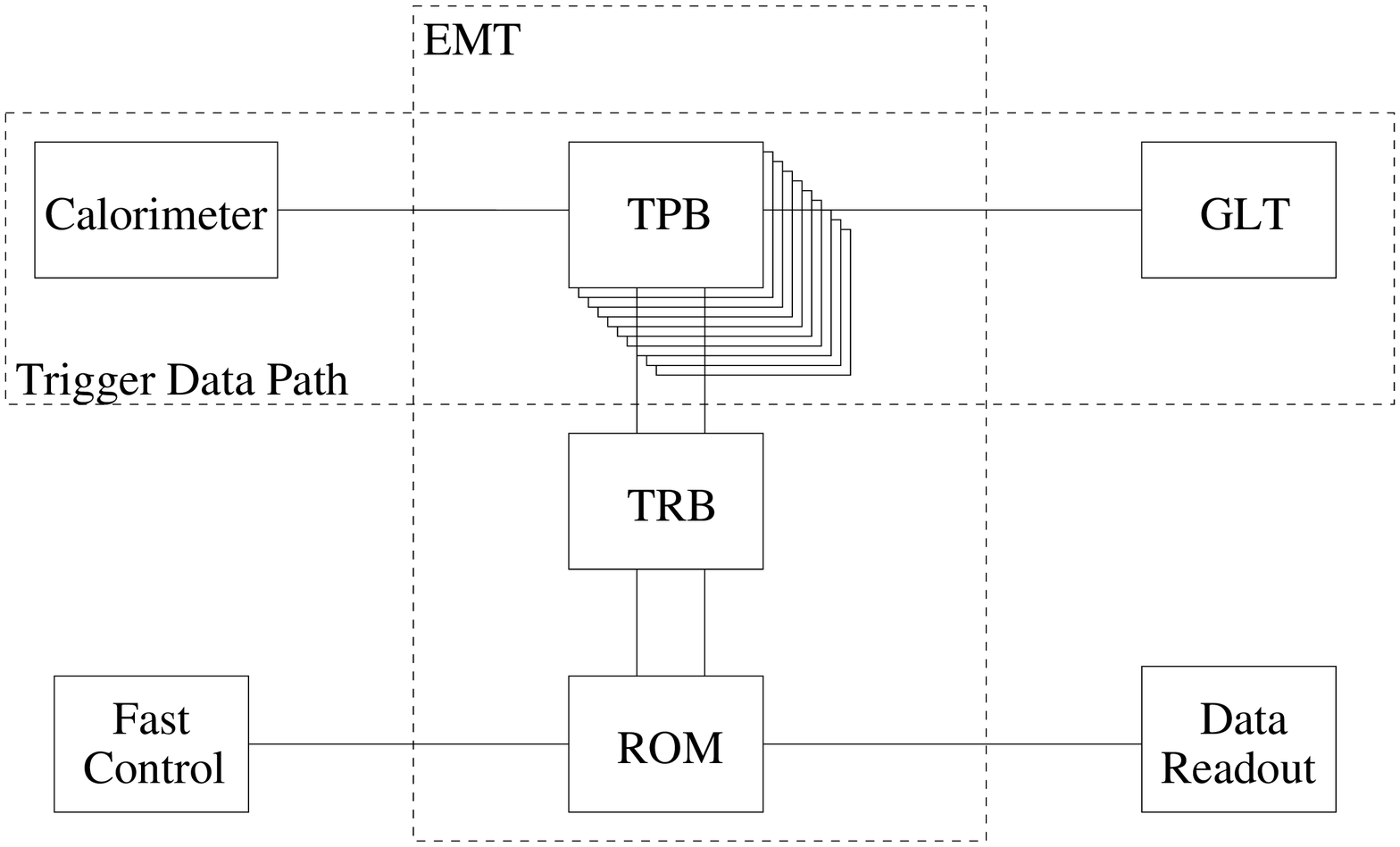}}
\caption{Simplified diagram of the relationships between the
boards in the EMT and the external systems. The calorimeter and GLT
data connections are on dedicated cables, the TRB to TPB
connections are across a custom backplane and the ROM
to TRB connections are via fibre-optic cable.}
\end{figure}

The TPB's are housed in a 9U VME crate with a standard J1 backplane and
three custom backplanes covering J2 and J3.
The EMT was built to operate with the same protocol as every front-end
system in \BaBar, in order to minimise the need for system-specific 
readout software.
This protocol is implemented via a TX/RX fibre-optic 
connection~\cite{ref:glink} from a
\BaBar\ ReadOut Module (ROM)~\cite{ref:daq}, which is common to
every system in the detector. 
A split J2 and J3 backplane divides the crate into two independent sections,
one occupied by
the ROM and a related fast control and
timing module, the other containing the EMT cards.

The whole EMT is timed and controlled via the fibre-optic connection
from the ROM. The fundamental 59.5 MHz clock signal and all configuration 
and control commands are sent along this fibre and distributed across 
the custom backplane to the TPB's
via an optical-to-electrical converter TRansition Board (TRB)
in the centre of the crate.
Because of the high clock speed, all the 
custom backplane signals were implemented as differential
100 $\Omega$ ECL point-to-point connections.
In addition,
care was taken to route the differential lines on the custom backplane
as closely as possible and equalise the signal return delay times between
the TRB and each TPB.
The command protocol is decoded on the TPB by a ``Fast Control''
Xilinx 4013E FPGA, which also 
performs all control functions on the board. All output data from
the TPB's are sent via the backplane to the TRB where
they are converted to optical signals. Hence, the fibre-optic connection
back to the ROM
is used to transmit the event data and to read back configuration data, 
as standard in \BaBar.

The major components on the TPB are shown in Fig.~5 and a photograph
of a production board is shown in Fig.~6.
Besides the algorithm calculation itself, the TPB performs several
other functions that are described in the following paragraphs.
\begin{figure}[htbp]
\label{figure:tpb}
\smallskip
\centerline{\includegraphics[width=3.4in,clip]{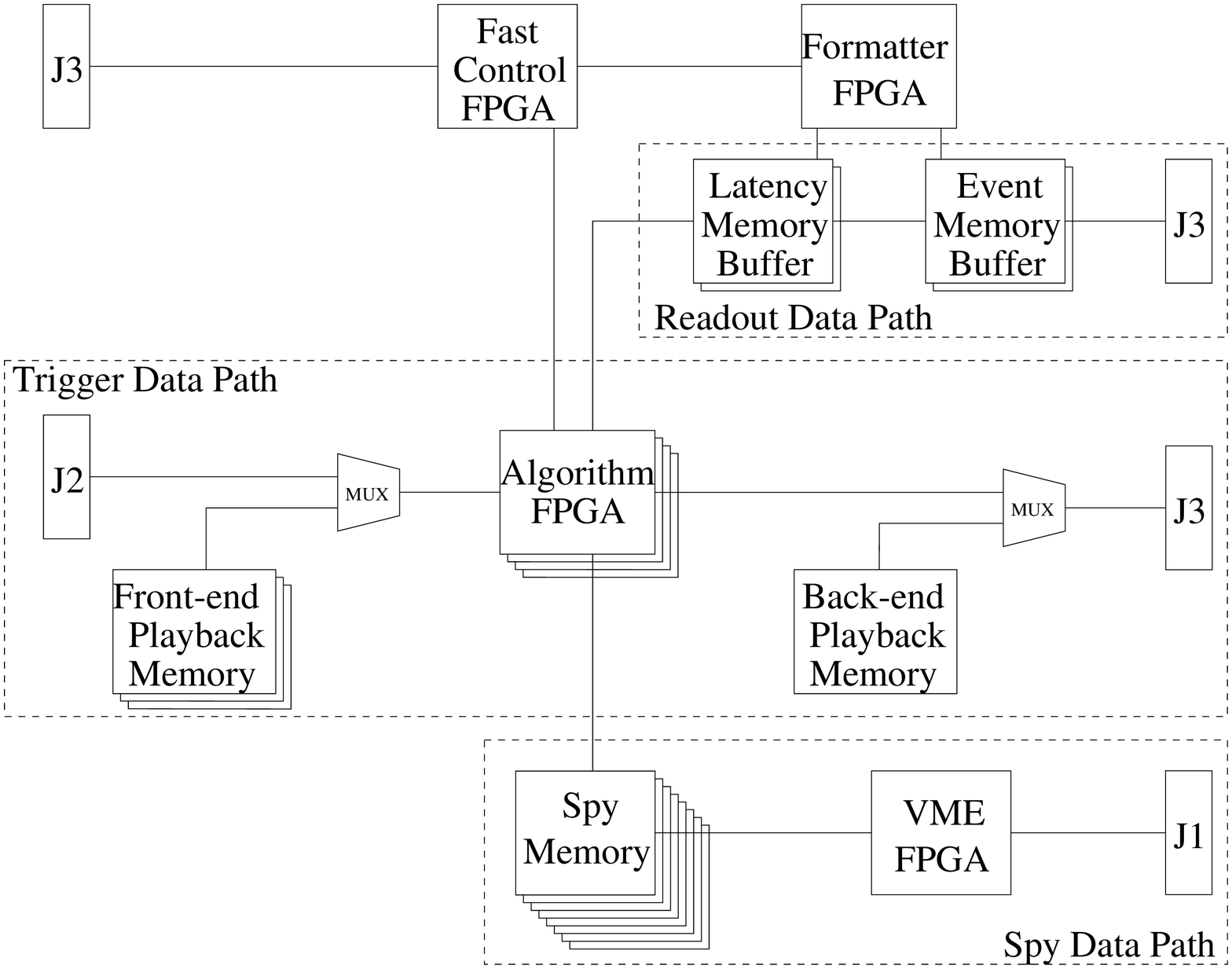}}
\smallskip
\caption{Major components on each TPB. See text for a description of the
function of each component.}
\end{figure}
\begin{figure}[htbp]
\label{figure:ProductionTpbTop}
\smallskip
\centerline{\includegraphics[width=3.4in,clip]{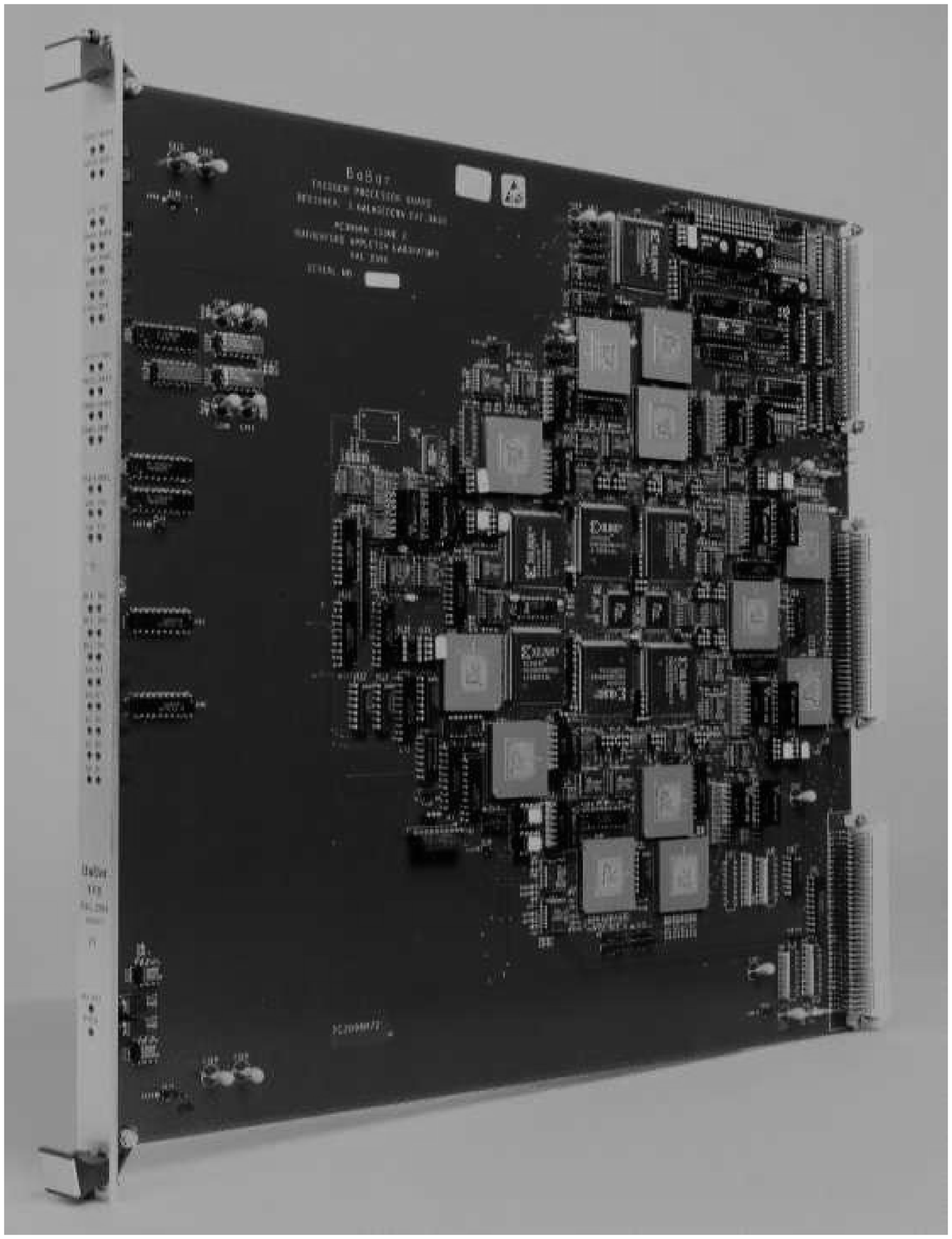}}
\smallskip
\caption{Photograph of the top side of a production TPB. The clustering of
the components is clearly visible. The board layout was optimised
to keep the data paths between the components as short as possible.}
\end{figure}

The data associated with the triggered event are read from
the EMT whenever there is a Level 1 trigger and are buffered in the TPB for
the 12 $\mu$s latency.
They are then stored in one of four event buffers until a readout request
is sent from the ROM. Each event buffer can store up to $\pm 2 \mu$s of data
around the event time, this length being a configurable parameter.
The latency and event buffers are configured and controlled by a ``Formatter''
Xilinx 4013E FPGA. These data are used offline for detailed checks on
the trigger performance and for tuning the algorithm configuration parameters.

The TPB's have large memory arrays which can be loaded with arbitrary
data patterns via the ROM. These come in two sets; the ``front-end'' and 
``back-end'' playback buffers. These can be configured to inject the
stored data to the front of the algorithm processor, instead of
the input data from the calorimeter, or to inject the stored data to
the EMT output, instead of the normal output data to the GLT.
Around 140~$\mu$s of data can be stored.
Any chosen data values can be clocked
through the entire EMT chain. This functionality
proved invaluable while testing the boards after
production.

In addition to the standard \BaBar\ control path, a read-only VME
interface was built into the TPB. This allows several on-board
``spy'' memories to be read at a low rate while the trigger is active.
These spy memories are filled with data from several locations in the
trigger algorithm data path, namely the raw input data, the
results of a number of intermediate processing stages and
the final output to the GLT. The memories allow around 140~$\mu$s of data to
be stored at a time. This allows detailed, bit-level validation that the
EMT is functioning correctly by comparing the spy data with a software model.
The combination of the spy and the playback memories significantly reduced
the time taken in the prototype and production testing cycles.

\section{Performance of the EMT}
The EMT was fully installed for the start of \BaBar\ data-taking in
May 1999 and has operated successfully thereafter.
Fig.~7 shows a typical event from the
perspective of the Level 1 trigger.
The detector is drawn end-on, with
the calorimeter information being shown
in the outermost layer. The solid blocks give the offline reconstructed
calorimeter 
information, the hashed blocks the EMT data. Four significant energy
deposits were identified, which together caused a trigger. In
addition, four tracks were found in the DCT and these are shown as the
lines in Fig.~7; they are clearly correlated with the calorimeter
deposits. The other DCT information, shown 
in the inner part of Fig.~7, indicates the track hits detected
(the small circles) overlayed on the
drift chamber cell geometry.
This event in fact satisfied
several redundant triggers, which is typical of the physics processes
of interest. Such events
allow cross-checks of the EMT and DCT trigger systems
against each other.
\begin{figure}[htbp]
\label{figure:L3Event}
\centerline{\includegraphics[width=3.4in,clip]{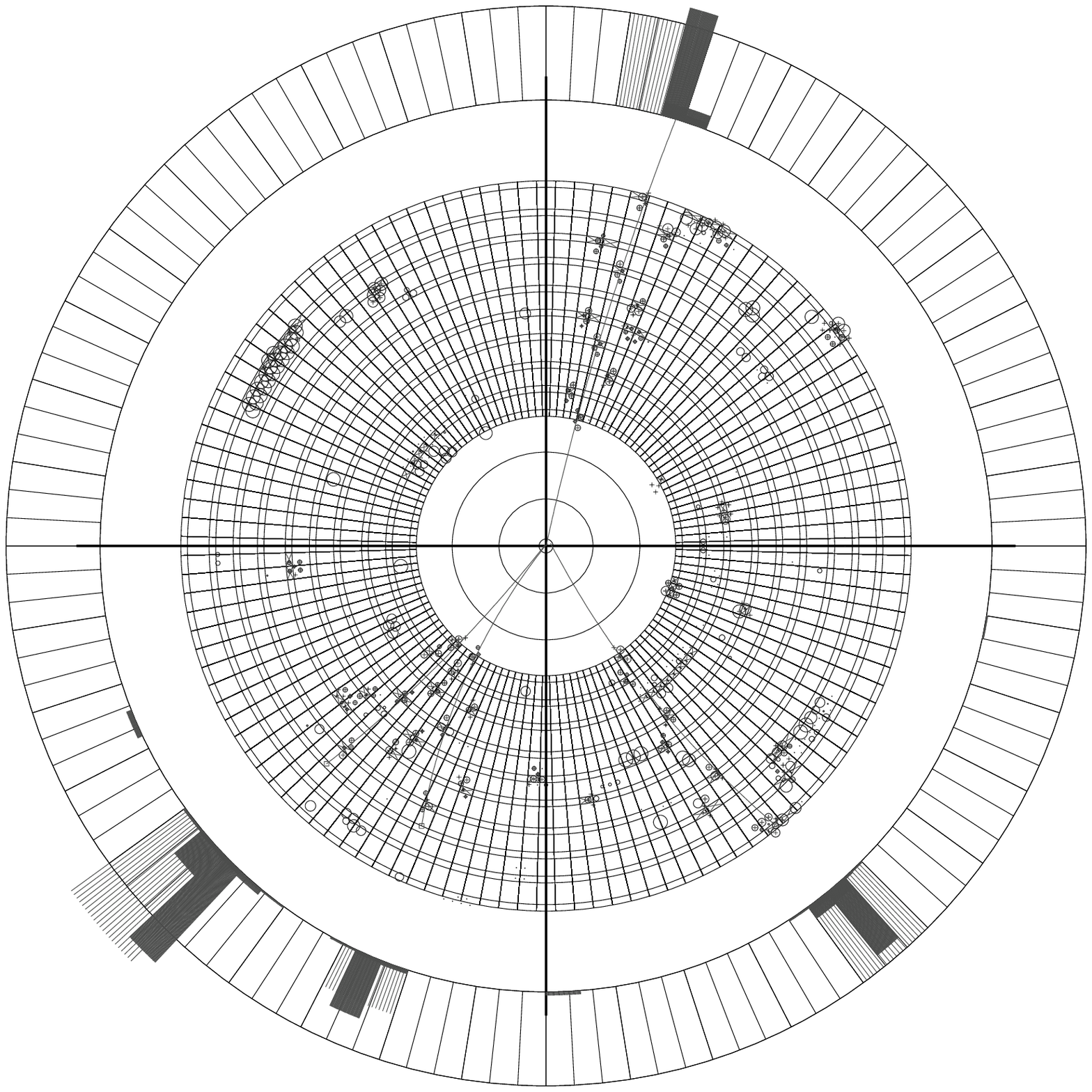}}
\caption{An example of a event which passed the Level 1 trigger.}
\end{figure}

No major components have needed to be replaced. The only minor
hardware concern has been the pin housing of the calorimeter-to-EMT
cables which was damaged through removal
of misplaced pins during the cable assembly.
The faulty housings have been replaced.

The EMT satisfied the technical requirements for latency and time jitter
without any tuning of the FIR filter weights beyond the values found from
a simulation study during the design phase. 
Fig.~8
shows the distribution of times determined in the EMT for individual
energy deposits, compared to a very much more accurate offline estimate
using the drift chamber data. This demonstrates the time of each deposit
is usually determined to much better than 1 $\mu$s. The overall requirement
for the Level 1 trigger of 1 $\mu$s applies to the whole event.
This event time is
determined in the GLT from all the primitives, 
potentially including those from the
DCT, and is effectively a weighted average.
The time resolution of the event is therefore
significantly better than that of each individual
energy deposit.
\begin{figure}[htbp]
\label{figure:jitter}
\centerline{\includegraphics[width=3.4in,clip]{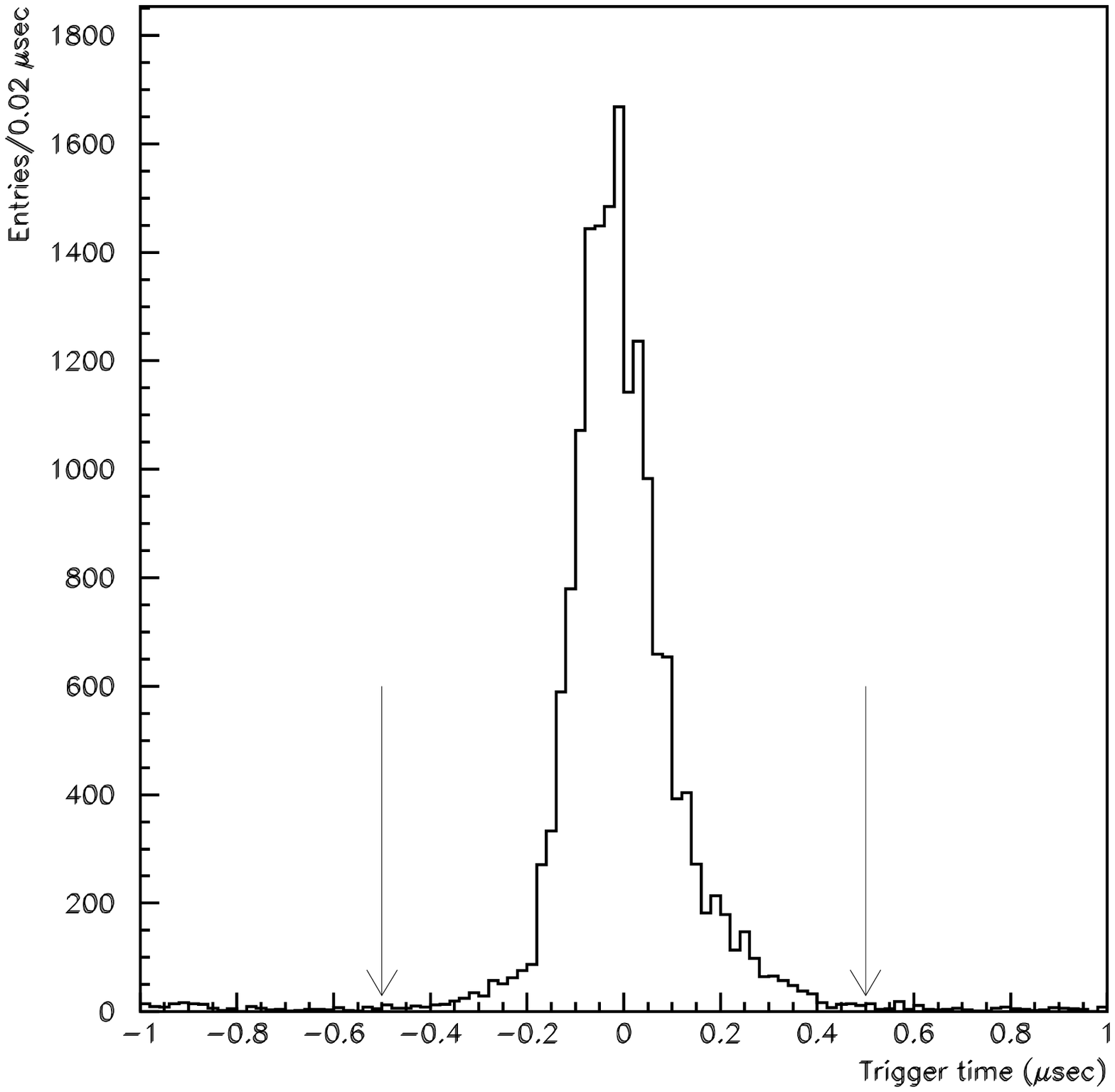}}
\caption{Distribution of energy deposit times determined in the EMT
compared with the time determined offline using the drift chamber, which is
accurate to a few ns. Deposits with energies above 120 MeV are used.
The arrows indicate the allowed 1 $\mu$s range.
}
\end{figure}

Typical Level~1 trigger
rates are below 1~kHz, well within the 2 kHz limit. 
Fig.~9
shows the overall rate for DCT and EMT triggers combined as 
a function of the PEP-II beam currents.
These rates
were achieved despite 
a factor of twenty times higher backgrounds
than expected levels, showing the importance of designing for high
background rates from the outset. 
The GLT uses a combination of EMT-only triggers, DCT-only triggers
and EMT-DCT combined triggers. In addition, an event often has multiple
trigger conditions satisfied simultaneously. It is therefore not possible
to give an unambiguous value for a trigger rate due to the EMT;
the rate of events which
only trigger because of the EMT is around 14\% of the total, while
removing the EMT completely from the Level~1 trigger would make the
rate fall by around 50\%.
\begin{figure}[htbp]
\label{figure:l1rate-vsher}
\centerline{\includegraphics[width=3.4in,clip]{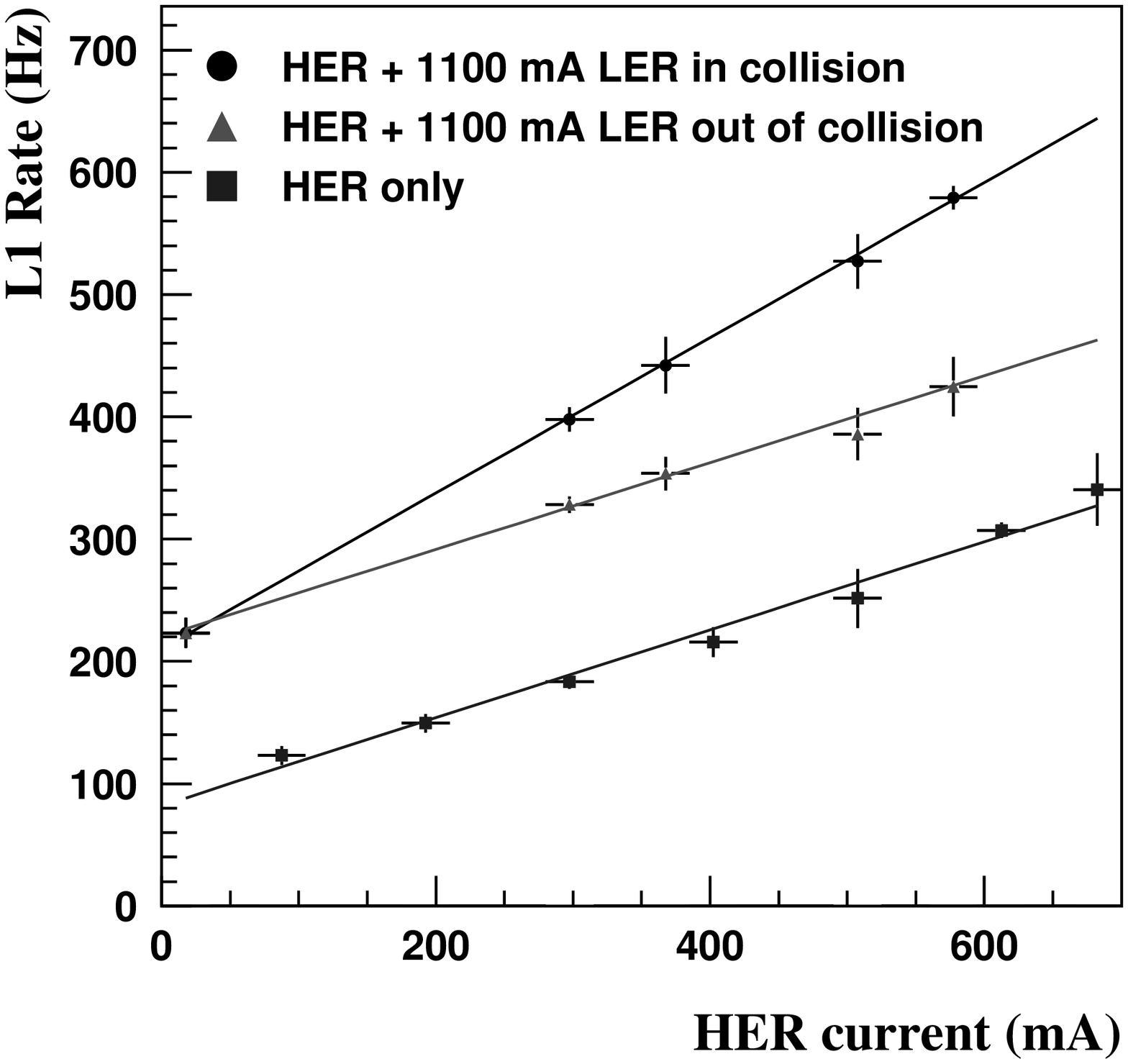}}
\caption{Typical Level~1 trigger rates as a function of the beam current
in the high energy ring (HER) of PEP-II.
The lowest line shows the
rate for beam in the HER only. The intermediate line shows the rate
with beam in the HER and 1100 mA of beam current in the low energy 
ring (LER), but with the beams being steered so as not to collide.
The highest line shows the same situation but with the HER and LER beams
colliding. This latter situation
corresponds to the normal operation of the PEP-II machine.
}
\end{figure}

The EMT efficiency for finding individual energy deposits above
threshold is above 99\%, except in regions with noisy channels 
which had to be excluded from the trigger (see below). 
Fig.~10
shows this
efficiency as a function of the deposited energy around the lowest
threshold used of 120~MeV. This threshold is designed to give a good
efficiency for minimum ionising particles, which deposit an average
of 180~MeV. It is seen that the trigger is at full efficiency by this
energy.
\begin{figure}[htbp]
\label{figure:mturnon}
\centerline{\includegraphics[width=3.4in,clip]{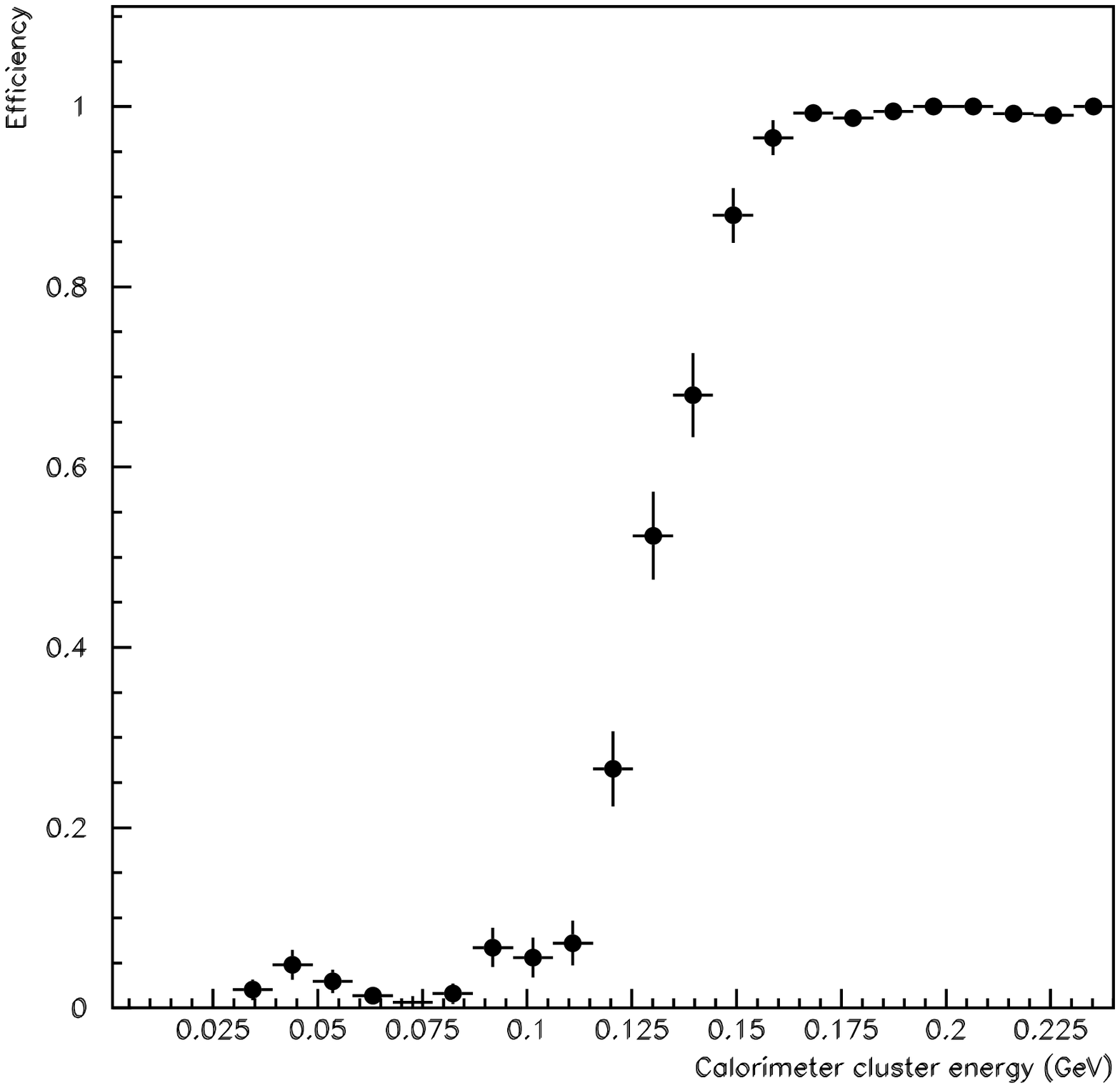}}
\caption{Efficiency for energy deposits close to the value of the
lowest threshold used.
}
\end{figure}

During the initial data-taking period, several faulty channels 
were found which sporadically became noisy. 
Whenever this occurred, it was necessary to 
modify a configurable mask to exclude these noisy regions from the
trigger. Some were due to isolated calorimeter channels which were
then disabled individually within the calorimeter electronics, allowing
the others in the 24 channel sum to continue to be used. However,
some noisy regions
were due to faults downstream of the sum, in which case
the input from all 24 had to be disabled.
The average number of input channels masked out over the
first year was 1.5\%. Simulation studies indicate that these 
dead regions caused 
a negligible loss of efficiency for $B$ meson decays.

The high efficiency per energy deposit translates directly into high
efficiencies for physics processes. The EMT efficiency for Bhabha
events, with both outgoing particles falling within the calorimeter
acceptance, has been measured to be at least 97\%, with some of the apparent
inefficiency being due to muon pairs contaminating the sample used to measure
the efficiency. The EMT efficiency for $B$ meson final states is more than
99\%, where the high figure is achieved through multiple redundancy;
typically the GLT requires four of the $\phi$ sums to be above the
120 MeV threshold, but hadronic events
have around ten such deposits on average.

\section{Conclusions}
The EMT has performed well during the first year of \BaBar\ data-taking.
The hardware has had no significant problems and the high-speed
FPGA design has functioned correctly throughout this period.
The trigger has proven to be robust against backgrounds
that were significantly higher than expected,
has been flexible enough to deal with
changing conditions, and has maintained a very high efficiency for all
physics collisions of interest.

\section{Acknowledgment}
The authors would like to thank Su Dong for his help
in the preparation of this paper.


\nocite{*}
\bibliographystyle{IEEE}

%

\end{document}